# Modulational instability of few cycle pulses in optical fibers

Amarendra K. Sarma*
Department of Physics, Indian Institute of Technology Guwahati, Guwahati-781039, Assam, India.
*Electronic address: aksarma@iitg.ernet.in

We investigate the modulation instability of a mathematical model appropriate for few cycle optical pulses with pulse duration as short as one carrier oscillation cycle in the context of a standard silica fiber operating at the telecommunication wavelength 1550 nm. Propagation of soliton-like few-cycle pulses in the medium is subject to the fulfillment of the modulation instability criteria.

Since the experimental generation of few cycle optical pulses with durations of the order of one attosecond, or pulses with the durations of only a few periods of optical radiation, research in the area of few cycle optical pulses have got a tremendous boost [1-2]. This is more so, owing to the astonishing possible applications of few cycle optical pulses in many diverse areas such as, ultrafast spectroscopy, metrology, medical diagnostics and imaging, optical communications, manipulation of chemical reaction and bond formation, material processing etc. [3-4]. In this context an appropriate mathematical model describing the dynamics and propagation of few cycle optical pulses in linear and nonlinear media have been researched by many authors [5-7]. This is mainly motivated by the fact that the so called Nonlinear Schrodinger Equation (NLSE), which is routinely used as the governing equation for describing pulse propagation in a media, is inadequate in the few cycle regimes. The fundamental reason attributed to the failure of NLSE in the few cycle regimes is due to the breakdown of the so called slowly varying envelope approximation (SVEA) [8-10]. Many authors have attempted to modify the SVEA so that it might be extended to the few cycle regimes. The first widely accepted model in this regard has been developed by Brabec and Krausz [5]. Some other authors have offered non-SVEA models also [11-13]. However, the model equation proposed by Brabec and Krausz have been used most extensively and successfully in various contexts [14-17].

In this work, we have studied the modulation instability (MI) of few cycle pulses in an optical fiber exhibiting an instantaneous third order cubic nonlinearity. The main motivation behind the work is that, the modulation instability is closely related to the existence of optical solitons in a nonlinear media like optical fiber [18]. And optical soliton effect may play a key role in the generation of few cycle pulses and their propagation through an optical fiber [19-20]. In passing, we would like to mention that modulation instability is a fundamental and ubiquitous process that appears in most nonlinear systems in nature [21-25]. The propagation of a few cycle optical pulses in a weakly dispersive nonlinear medium displaying instantaneous third order cubic nonlinearity is governed by the following equation [5, 26]:

$$\left[\nabla_{\perp}^2 + 2i\beta_0 \frac{\partial}{\partial z}\left(1 + \frac{i}{\omega_0}\frac{\partial}{\partial t}\right) - \beta_0 \beta_2 \frac{\partial^2}{\partial t^2}\left(1 + \frac{i}{\omega_0}\frac{\partial}{\partial t}\right)\right] A(\mathbf{r},t)$$
$$= -\frac{3\omega_0^2 \chi^{(3)}(\omega_0)}{c^2}\left(1 + \frac{i}{\omega_0}\frac{\partial}{\partial t}\right)^2 |A(\mathbf{r},t)|^2 A(\mathbf{r},t) \quad (1)$$

where $A(\mathbf{r},t)$ is the pulse envelope centered at frequency $\omega_0$ propagating along the $z$-axis, $t$ is the retarded time, $\beta_2$ is the group velocity dispersion (GVD) parameter, $\chi^{(3)}(\omega_0)$ is the third order susceptibility evaluated at $\omega_0$ and $\nabla_\perp^2 = \partial^2/\partial x^2 + \partial^2/\partial y^2$ is the diffraction operator. $\beta_0 = \omega_0 n/c$ is the wave vector where $n$ is the refractive index and $c$ is the speed of light in free space. For an optical fiber, neglecting the diffraction term and writing $(1+i/\omega_0 \partial/\partial t)^2 \approx (1+2i/\omega_0 \partial/\partial t)$, we may write Eq. (1) in the following form:

$$\frac{\partial A}{\partial z} + \frac{i}{\omega_0}\frac{\partial^2 A}{\partial z \partial t} + \frac{i}{2}\beta_2 \frac{\partial^2 A}{\partial t^2} - \frac{\beta_2}{2\omega_0}\frac{\partial^3 A}{\partial t^3} = i\gamma|A|^2 A - \frac{2}{\omega_0}\gamma \frac{\partial}{\partial t}\left(|A|^2 A\right) \qquad (2)$$

where $\gamma = 3\beta_0 \chi^{(3)}(\omega_0)/2n^2$ is the nonlinear parameter with $n$ as the refractive index of the core of the fiber medium. For ease of our analysis we write Eq. (2) in the normalized units as follows:

$$\frac{\partial u}{\partial \xi} + is\frac{\partial^2 u}{\partial \xi \partial \tau} + \frac{i}{2}\delta \frac{\partial^2 u}{\partial \tau^2} - \frac{1}{2}\delta s \frac{\partial^3 u}{\partial \tau^3} - i|u|^2 u + 2s\frac{\partial}{\partial \tau}\left(|u|^2 u\right) = 0 \qquad (3)$$

where $u$ is the normalized amplitude and

$$\xi = z/L_D, \tau = t/T_0, L_D = T_0^2/|\beta_2|, A = \sqrt{P_0}Nu, N = \sqrt{\gamma P_0 L_D}, s = 1/\omega_0 T_0 \qquad (4)$$

in which $\xi$ and $\tau$ are the normalized propagation distance and time respectively, $P_0$ is the peak power of the incident pulse, $L_D$ is the dispersion length, $N$ is the so called soliton order [18] and $s$ is the self-steepening (SS) parameter. $\delta = \text{sgn}(\beta_2)$. It should be noted that the novelty of Eq.(3) lies in the presence of the second term which refers to space-time coupling and the fourth term which couples the third order derivative of the pulse envelope with the self-steepening and the GVD parameter. On the basis of Eq. (3) we would now investigate the MI of few cycle pulses. Eq. (3) has a steady state solution given by $u = u_0 \exp\left[iu_0^2 \xi\right]$, where $u_0$ is the constant amplitude of the incident plane wave. We now introduce perturbation $a(\xi,\tau)$ together with the steady state solution to Eq.(3) and linearize in $a(\xi,\tau)$ to obtain

$$\frac{\partial a}{\partial \xi} - iu_0^2(a+a^*) + su_0^2\left(3\frac{\partial a}{\partial \tau} + 2\frac{\partial a^*}{\partial \tau}\right) + is\frac{\partial^2 a}{\partial \xi \partial \tau} + \frac{1}{2}\delta\left[i\frac{\partial^2 a}{\partial \tau^2} - s\frac{\partial^3 a}{\partial \tau^3}\right] = 0 \qquad (5)$$

Separating the perturbation to real and imaginary parts, according to $a = a_1 + ia_2$, and assuming $a_1, a_2 \propto \exp\left[i(K\xi - \Omega\tau)\right]$, where $K$ and $\Omega$ are the wave number and the frequency of perturbation respectively, from Eq.(5) we obtain the following dispersion relation

$$K = \left[\left(-2\Omega s u_0^2 \pm \tfrac{1}{2}\sqrt{P-RQ}\right)/R\right] \qquad (6)$$

where $P = 16s^2 u_0^4 \Omega^2, Q = \Omega^4 - \Omega^6 s^2 + 4\delta u_0^2 \Omega^2 - 12\delta\Omega^4 s^2 u_0^2 - 20s^2 \Omega^2 u_0^4$ and $R = \Omega^2 s^2 - 1$. Clearly, from Eq.(6), we observe that modulation instability exists only if $RQ - P > 0$ and $R \neq 0$. Under these MI conditions we obtain the gain spectrum $g(\Omega)$ of the modulation instability as follows

$$g(\Omega) = 2\text{Im}(K) = 2[RQ - P]^{\frac{1}{2}} \qquad (7)$$

We note that in the absence of self-steepening, i.e. for $s = 0$, the gain becomes maximum at $\Omega_{\max} = \pm\left[-2\delta u_0^2\right]^{\frac{1}{2}}$ which clearly indicates that we must have $\delta = -1$ or, in other words, MI is possible in optical fiber only in the anomalous dispersion regime, an already well-established result in nonlinear fiber optics [18]. Now in order to assess the role of pulse width of few cycle pulses on modulation instability in a standard silica fiber operating at 1550 nm, in Fig. 1 we depict the gain spectra as a function of the normalized frequency for $T_0 = 10, 7, 5$ and $3$ fs which corresponds to $s = 0.08, 0.12, 0.16$ and $0.27$ respectively, with initial normalized pulse width $u_0 = 1$. It is important to note that a standard silica fiber, used in telecommunications, operating at 1550 nm usually have the following typical parameters [18]: $\beta_2 = -20 \, \text{ps}^2/\text{km}$, nonlinear index $n_2 = 2.6 \times 10^{-20} \, \text{m}^2/\text{W}$.

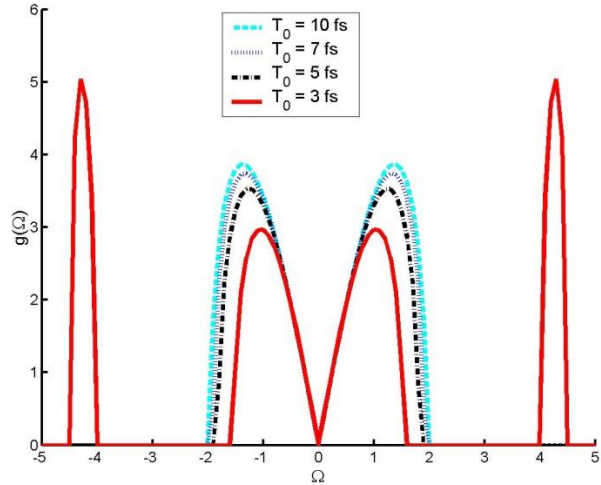

Fig. 1. (Color online) Modulational instability gain as a function of normalized frequency for four different values of pulse width with $u_0 = 1$ in a standard silica fiber operating at the telecommunication wavelength 1550 nm.

It can be clearly seen that the gain spectrum is symmetric with respect to $\Omega = 0$. We observe from Fig. 1 that for the given input power, the modulation instability gain decreases with decrease in the pulse width or equivalently with increase of the self-steepening parameter.

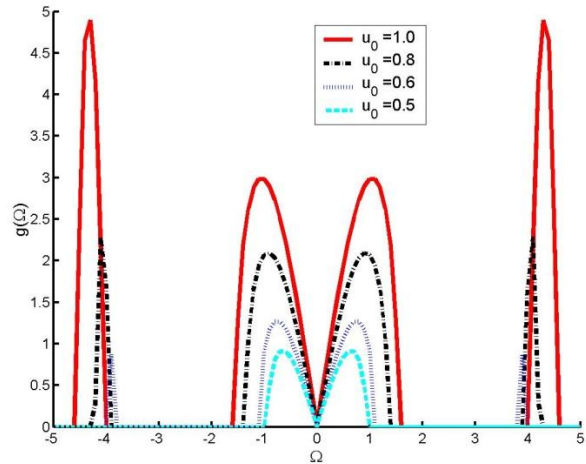

Fig. 2. (Color online) Modulational instability gain as a function of normalized frequency for different values of the initial normalized amplitude for a few cycle pulse with $T_0$=3 fs in a standard silica fiber operating at the telecommunication wavelength 1550 nm.

Physically speaking, if a probe wave at a frequency $\omega_0 + \Omega$ were to propagate with the CW beam at $\omega_0$, it would experience a net power gain given by Eq.(7) as long as $RQ - P > 0$. Eventually due to MI gain, the CW beam would break up spontaneously into a periodic pulse train known as solitons. These soliton-like pulses exist whenever the conditions $RQ - P > 0$ and $R \neq 0$ are satisfied. The appearance of the sidebands located around $\Omega = 0$ is the clear evidence of modulation instability. An interesting feature appears when the pulse width approaches 3.5 fs or less. For example when $T_0$=3 fs, we find that along with the usual peaks, two side band peaks appear at higher frequencies. However, as evident from Fig.2, where we plot the modulation instability gain as a function normalized perturbation frequency $\Omega$ for different values of normalized amplitude $u_0$ for a given pulse width, say $T_0 = 3$ fs, these side band peaks vanish as the input peak power is reduced beyond a certain value, $u_0 = 0.5$ for the given example here. We also note that the side band peaks move towards the centre as the initial amplitude or peak power is reduced. It may be relevant to mention here that in order to study the exact nature of the soliton-like pulses we need to solve Eq. (3) analytically or numerically. To have an idea about the evolution of such a pulse we solve Eq. (3) numerically in the anomalous dispersion regime by the so called split-step Fourier method [18], considering the input to be $u(0, \tau) = \sec h(\tau)$. For the ease of numerical calculations we have neglected the space-time coupling term in Eq. (3). The input and output profile of the envelope of a pulse with $T_0 = 3$ fs, after propagation of a distance around $153 \ \mu$m (in real units), is depicted in Fig. (3).

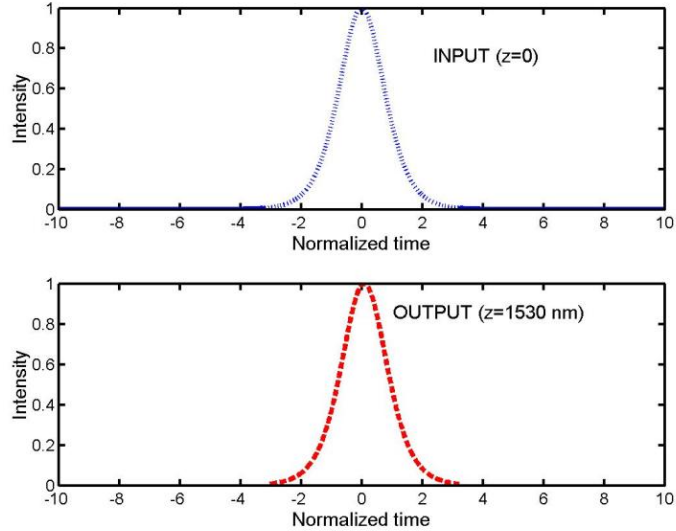

Fig. 3. (Color online) Input and output temporal profile of the pulse envelope of a few cycle pulse with $T_0$=3 fs. The output pulse refers to a pulse at a distance of 1530 nm.

We observe that the pulse is maintaining its solitonic character during its propagation. For more rigorous study on the dynamics of few cycle pulses in various contexts, where the existence of solitons are assumed a priori, readers are referred to Ref.[13-15, 19, 20].

To conclude, the modulation instability of a mathematical model appropriate for few cycle optical pulses with pulse duration as short as one carrier oscillation cycle have been investigated in the context of a standard silica fiber operating at the wavelength 1550 nm. A nonlinear dispersion relation is worked out using standard methods. We have found that the growth spectrum of MI is sensitive to the pulse width for a given input power. We have also found that by varying the input peak power of the pulse one might control the peaks of the side bands of the MI gain spectrum of a few cycle optical pulse with a given initial pulse width. The existence of MI clearly indicates that the soliton concept is still relevant even in the attosecond regime. Soliton-like few-cycle pulses are supported in the medium provided the modulation instability criterion is satisfied.

The author sincerely thanks the Department of Science and Technology, Government of India for a research grant (grant no. SR/FTP/PS-17/2008).

-------------------------------------------------------------------------------------------------------------